# Possible spin gapless semiconductor type behaviour in CoFeMnSi epitaxial thin films

Varun K. Kushwaha, Jyoti Rani, Ashwin Tulapurkar, and C. V. Tomy

# Possible spin gapless semiconductor type behaviour in CoFeMnSi epitaxial thin films

Varun K. Kushwaha,[1,a)] Jyoti Rani,[1] Ashwin Tulapurkar,[2] and C. V. Tomy[1]

[1]Department of Physics, Indian Institute of Technology Bombay, Mumbai 400076, India
[2]Department of Electrical Engineering, Indian Institute of Technology Bombay, Mumbai 400076, India

Spin-gapless semiconductors with their unique band structure have recently attracted much attention due to their interesting transport properties that can be utilized in spintronics applications. We have deposited the thin films of a quaternary spin-gapless semiconductor CoFeMnSi Heusler alloy on MgO (001) substrates using a pulsed laser deposition system. These films show epitaxial growth along the (001) direction and display a uniform and smooth crystalline surface. The magnetic properties reveal that the film is ferromagnetically soft along the in-plane direction and its Curie temperature is well above 400 K. The electrical conductivity of the film is low and exhibits a nearly temperature independent semiconducting behaviour. The estimated temperature coefficient of resistivity for the film is $-7 \times 10^{-10} \ \Omega \ m/K$, which is comparable to the values reported for spin-gapless semiconductors. *Published by AIP Publishing.* https://doi.org/10.1063/1.4996639

In recent years, spin-gapless semiconductors (SGSs),[1–3] because of their unique band structure, have attracted much attention in the fields of condensed matter physics and materials science due to their potential applications in spintronics. SGSs provide an intermediate bridge between the half-metallic ferromagnets and zero bandgap semiconductors. These materials exhibit a bandgap in one spin channel ($\downarrow$) and zero bandgap in the other spin channel ($\uparrow$). This remarkable combination of electronic band properties in SGSs leads to several useful and unique transport properties, even at room temperature. Some of these properties include[1] the requirement of zero threshold energy to excite the electrons from the valence band to the conduction band, 100% spin polarization of the excited charge carriers (electrons and holes), and ability to switch between n-type or p-type spin-polarized carriers by changing the sign of the applied gate voltage. All these properties were predicted[1] to be extremely sensitive to external influences such as pressure, electric field, magnetic field, electromagnetic radiation, and impurities. SGS properties have been observed in Heusler alloys ($Mn_2CoAl$,[4] CoFeMnSi (CFMS),[5,6] CoFeCrAl,[5] Ti$_2$MnAl,[7] etc.), Co-doped PbPdO$_2$,[8] strained zigzag silicene nanoribbons,[8] and strained Ti$_2$C monolayers.[9] Recently, a few equiatomic quaternary Heusler alloys (EQHAs),[10,11] XX′YZ (where X, X′, and Y are transition metals and Z is a main group element with 1:1:1:1 stoichiometry), which show SGS behaviour, have attracted much interest due to their exceptional magnetic and transport properties over half-metallic ferromagnets (HMFs). EQHAs crystallize in the Y-type or LiMgPdSn prototype structure with space group F4̄3m (#216).[5,6,10–14] The title alloy, CoFeMnSi (CFMS), is an EQHA that had originally been reported[12–14] as a half-metallic ferromagnet (HMF) on the basis of magnetic and hard X-ray spectroscopy measurements (these measurements in general cannot predict the metallic or semiconducting nature of the material). Later, this alloy was identified to be a SGS theoretically,[5] which was further confirmed experimentally using electrical conductivity (semiconducting behaviour) and Hall measurements.[6] However, for spintronics applications, the unique properties of SGS have to be realized in thin films. There are only three SGS Heusler alloys that have been prepared in the film form, e.g., Mn$_2$CoAl,[15–17] CoFeCrAl,[18] and Ti$_2$MnAl.[19] Here, we report the fabrication of the epitaxial thin film of yet another SGS alloy, CFMS, followed by the structural, magnetic, and transport properties. From the structure, surface morphology, and surface topography analyses, we have shown that the films grown are epitaxial and homogeneous with the surface roughness of the order of 0.7 nm. Magnetic measurements confirm the ferromagnetic nature of the film, and the resistivity measurements indicate the possibility of a spin gapless semiconductor.

Thin films of CFMS were grown on single crystal substrates of MgO (001) using an excimer KrF ($\lambda = 248$ nm) pulsed laser deposition (PLD) system. The target to substrate distance was 4 cm, the laser repetition frequency was 2 Hz, and the laser fluence was fixed at 3.5 J/cm$^2$. Prior to the thin film deposition, the MgO substrate was first annealed at 800 °C (base pressure $\sim 2.0 \times 10^{-6}$ mbar) for nearly 1 h, the temperature was brought down to 500 °C, and then the CFMS thin film was deposited (we have finalized a substrate temperature 500 °C as the optimum substrate temperature for obtaining the best quality thin films, after studying the growth conditions at various substrate temperatures). The deposited films were annealed *in-situ* at 700 °C for 30 min to enhance the crystallization and chemical ordering.

The structure of the CFMS thin films and their thickness were analysed by X-ray diffraction (XRD) and X-ray reflectivity (XRR), respectively, using a high resolution X-ray diffractometer with Cu-K$_\alpha$ radiation ($\lambda = 1.5406$ Å). The surface morphology and the topography of the films were determined by using a field emission gun-scanning electron microscope (FEG-SEM) and an atomic force microscope (AFM), respectively. The elemental composition in the film was determined using energy dispersive X-ray spectroscopy (EDX) attached to FEG-SEM. The temperature (T) and the magnetic field (H)

a)Author to whom correspondence should be addressed: varun@phy.iitb.ac.in

dependence of magnetization (M) were carried out using a superconducting quantum interference device-vibrating sample magnetometer (SQUID-VSM, Quantum Design, USA). The physical property measurement system (PPMS, Quantum Design, USA) was used to measure the electrical resistivity ($\rho_{xx}$) and the Hall resistivity ($\rho_{xy}$) as a function of temperature and magnetic field using the four-probe and five-probe methods, respectively.

Figure 1(a) shows the room temperature $\omega$-$2\theta$ (out-of-plane) X-ray diffraction pattern of one of CFMS thin films deposited on the MgO (001) substrate. It is clear from the figure that the dominant diffraction peaks are (002) and (004) from CFMS, in addition to the peaks from the MgO substrate, indicating the epitaxial growth of the CFMS film along the (001) orientation. The lattice constant (a) of the CFMS film was estimated to be around 5.646 Å, which compares well with the reported experimental value[6,12,14] for bulk samples. The thickness of the CFMS thin film (~18 nm) was calculated from the XRR spectra [see Fig. 1(b) and its inset] using the modified Bragg equation:[20]

$$\theta_m^2 = \theta_c^2 + \left(\frac{\lambda}{2t}\right)^2 m^2 \tag{1}$$

where $m$ is an integer (the index for each oscillation maxima, $\lambda$ is the wavelength of the X-ray, $t$ is the thickness of the thin film, $\theta_c$ is the critical angle (in radians) for total reflection, and $\theta_m$ is the Bragg angle (in radians) of the $m^{th}$ oscillation maxima.

In order to confirm the epitaxial nature of the thin film, we show the $\phi$-scan patterns for the (220) peak ($2\theta = 45.3°$, $\chi = 45°$) and the (111) peak ($2\theta = 27.3°$, $\chi = 54.7°$) in Figs. 1(c) and 1(d), respectively. For the (220) $\phi$-scan, we observe four well-defined peaks, periodically separated from one another by an angular difference of 90° [see Fig. 1(c)] indicating a four-fold symmetry in the sample plane. The (111)

$\phi$-scan also displays a four-fold symmetry [see Fig. 1(d)], indicating either a Y-type structure or an L2$_1$ structure[21] for our CFMS film. In the absence of some specialized measurements like neutron diffraction, it is very difficult to distinguish between the L2$_1$ and Y-type structures since the scattering factors for Co and Fe are almost identical for X-ray (Cu-K$\alpha$ radiation). Further, each peak in the (220) $\phi$-scan is separated by an angle of 45° to the corresponding peak in the (111) $\phi$-scan, which is expected for the cubic crystal structure of the CFMS alloy. All these results confirm the epitaxial growth of the CFMS thin film on the MgO substrate.

The order parameters for the CFMS film were calculated using the extended Webster model.[22] The degree of Y-type and B2 ordering can be calculated from the long-range order parameters

$$S_{B2}^2 = \frac{I_{002} \cdot I_{004}^f}{I_{004} \cdot I_{002}^f} \quad \text{and} \quad \left(\frac{S_Y(3 - S_{B2})}{2}\right)^2 = \frac{I_{111} \cdot I_{220}^f}{I_{220} \cdot I_{111}^f}, \tag{2}$$

where $I_{hkl}$ is the experimental intensity from X-ray diffraction and $I_{hkl}^f$ is the theoretical diffraction intensity calculated for a fully ordered Y-type CFMS alloy. [Note: in Eq. (2), we have written $S_Y$ instead of $S_{L2_1}$ because EQHAs crystallize in the Y-type Heusler structure].[5,6,10–14,21] The values for $S_Y$ and $S_{B2}$ were found to be 94% and 95%, respectively, which indicate a high degree of Heusler-type order.

The surface morphology of the film obtained from the FEG-SEM image is shown in Fig. 2(a), which indicates that the film is homogeneous and consists of tiny crystal grains. The atomic ratio (%) of each element of the film determined by EDX is 24 (±2), which is very close to ideal stoichiometric composition for EQHA. In the absence of any impurity peaks in X-ray data, the obtained stoichiometry can be taken close to the ideal stoichiometry. Figures 2(b) and 2(c) show the surface topography of the thin film obtained from the AFM scans for a $2\,\mu m \times 2\,\mu m$ area of the thin film at room temperature in 2 dimensional and 3 dimensional spaces, respectively. The average roughness ($R_a$) obtained for this film is around 0.7 nm.

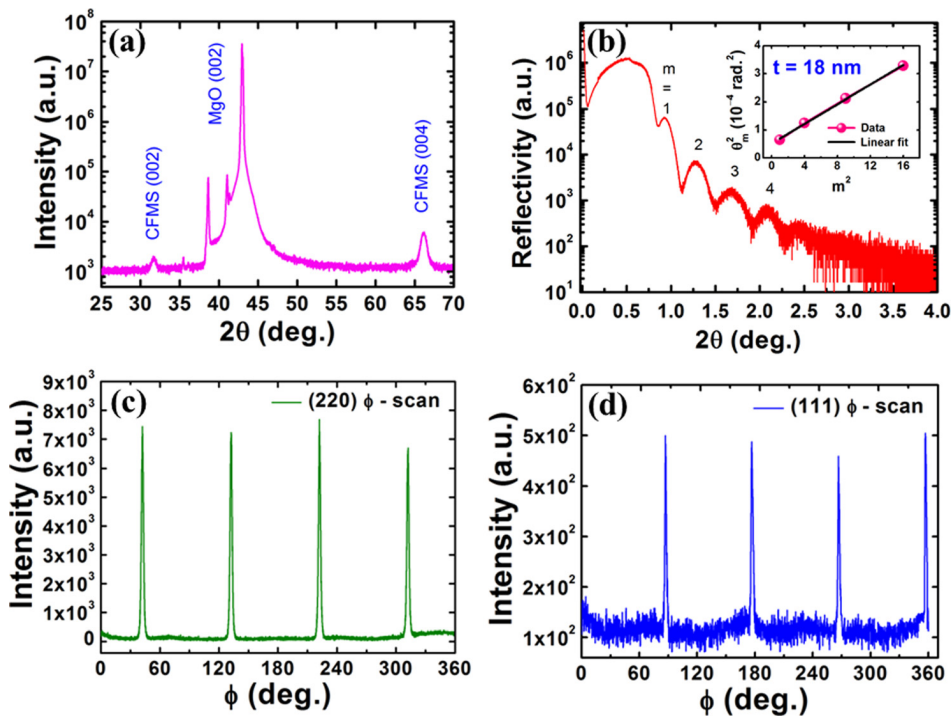

FIG. 1. (a) XRD $\omega$-$2\theta$ (out-of-plane) scan for the CFMS thin film on the MgO substrate. (b) XRR data of the film. (Inset: linear fit for calculating the film thickness.) $\phi$-scan of (c) (220) reflection and (d) (111) reflection.

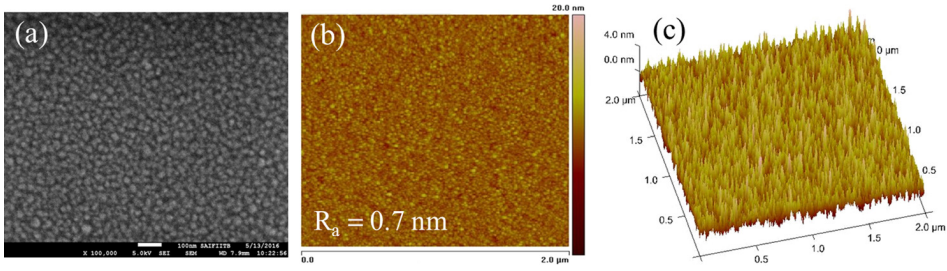

FIG. 2. (a) FEG-SEM image of the film. AFM image of the thin film surface over $2\,\mu m \times 2\,\mu m$ area: (b) two dimensional and (c) three dimensional space.

In order to corroborate the magnetic properties of the prepared thin film, we have measured the magnetization of the film as a function of temperature. The upper inset of Fig. 3 shows the temperature dependence of the magnetization (M-T) performed in zero field cooled (ZFC) and field cooled (FC) modes from 5 to 400 K in an applied magnetic field of 0.05 T. It is clear from the figure that the film is ferromagnetic and the Curie temperature ($T_C$) is well above 400 K ($T_C = 623$ K is reported for the bulk CFMS alloy[14]). The main panel of Fig. 3 displays the in-plane (IP) and out-of-plane (OP) isothermal magnetization (M-H) of the film as a function of applied magnetic field (H) at 300 K. The M-H loops indicate a clear anisotropy; the sample is ferromagnetically soft along the in-plane direction with a coercivity ($H_c$) of $\sim$0.03 T. The saturation magnetization ($M_s$) was found to be 3.28 $\mu_B$/f.u. at 300 K, which is less than the reported bulk value (3.7 $\mu_B$/f.u.).[14] Assumption of the same thickness as well as the usage of theoretical density ($\rho = 7.26$ g/cm$^3$) for calculating the mass of the thin film may also lead to the observed difference. During the preparation of this manuscript, we have come across a work reported by Bainsla *et al.*[23] where they have reported $M_s$ values from 2.15 $\mu_B$/f.u. to 3.54 $\mu_B$/f.u. at 300 K for films grown at different annealing temperatures using the sputtering technique. $M_s$ values obtained for higher annealing temperatures are comparable to the values we have obtained. The saturation magnetization as well as the coercivity does not show much of a temperature variation, as it is clear from the lower inset of Fig. 3, where the M-H curves at different temperatures are shown. Figure 4 shows the temperature dependence of electrical resistivity $\rho_{xx}(T)$ of the thin film in the temperature range of 2–300 K. The electrical resistivity decreases almost linearly with an increase in temperature, which clearly indicates a semiconducting behaviour. The temperature coefficient of resistivity (TCR) of the sample is $-7 \times 10^{-10}$ Ω m/K. This negative TCR value is much less than the TCR values of classical semiconductors[24] (e.g., TCR of Si is $-7.0 \times 10^{-2}$ Ω m/K). In the Heusler family of alloys, the two spin gapless semiconductors, Mn$_2$CoAl[4] and CoFeCrAl,[25] are also known to exhibit linear resistivity behaviour with negative TCR values of $-1.4 \times 10^{-9}$ Ω m/K and $-5 \times 10^{-9}$ Ω m/K, respectively, which are comparable with the TCR value of our sample. The electrical conductivity $\sigma_{xx}(T)$ value of our sample is around $2.86 \times 10^3$ S/cm at 300 K, which is very close to the $\sigma_{xx}(T)$ value of $2.44 \times 10^3$ S/cm for the spin gapless semiconductor, Mn$_2$CoAl,[4] and about two orders of magnitude lower than the half metallic ferromagnetic alloy Co$_2$MnSi ($\sim 10^5$ S/cm).[26] In order to estimate the charge-carrier concentration as a function of temperature, we have measured the Hall resistivity ($\rho_{xy} = R_O H + R_{AHE} M$, where $R_O$ and $R_{AHE}$ are the ordinary and the anomalous Hall coefficients, respectively, and $M$ is the magnetization at the given magnetic field) as a function of magnetic field at different temperatures. The charge-carrier concentration obtained from the ordinary Hall coefficient ($n = 1/e|R_O|$) at 300 K has a value of $8.9 \pm 0.1 \times 10^{20}$ cm$^{-3}$ and is nearly temperature independent down to 2 K (see the inset of Fig. 4). This carrier concentration value is two orders of magnitude less than that for typical metallic systems like Cu, Au, and Ag.[27] Our measurements also reveal the sign of $R_O$ to be positive, which indicates the presence of p-type charge carriers.

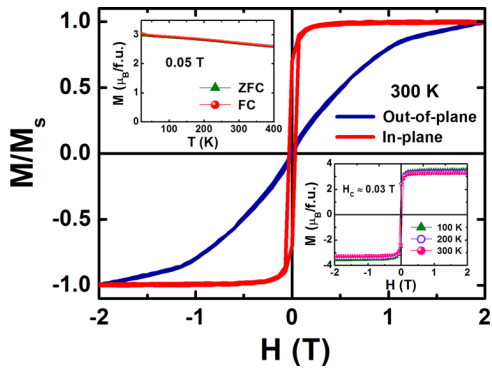

FIG. 3. The in-plane (IP) and out-of-plane (OP) normalized isothermal magnetization (M-H) curves of the film as a function of applied magnetic field (H) at 300 K. The upper inset shows the temperature dependence magnetization (M-T) curve and the lower inset shows the field dependence of magnetization (M-H) curve measured in in-plane (IP) at T = 100, 200, and 300 K.

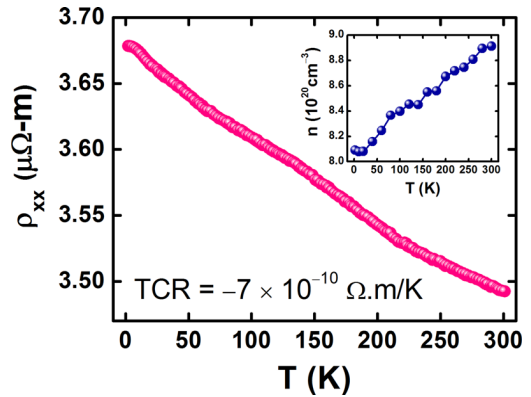

FIG. 4. Electrical resistivity of the CFMS thin film measured as a function of temperature. The inset shows the variation of the charge-carrier concentration as a function of temperature.

In conclusion, an epitaxial CFMS thin film (18 nm) was grown on a (001) oriented MgO single crystal substrate with a uniform and smooth crystalline surface using the pulsed laser deposition system. We have investigated the structural, magnetic, and transport properties of the CFMS thin film. Magnetization measurements reveal that the film is ferromagnetically soft along the in-plane direction and its Curie temperature is well above 400 K. The electrical resistivity of the film indicates a semiconducting behaviour with a TCR value of $-7 \times 10^{-10} \ \Omega \, m/K$ which is comparable with the TCR values reported for SGS Heusler alloys and is much less than the TCR values of classical semiconductors. Nearly temperature independent and low electrical conductivity of the order of $2.86 \times 10^3$ S/cm indicates a possible SGS behaviour and is in agreement with experimentally reported SGS behaviour. Point-contact Andreev reflection (PCAR) measurements will be useful to measure the spin-polarization in these CFMS thin films and to compare with the bulk samples. One of the main impacts of the SGS materials is that they can be used as the SGS/semiconductor heterostructures to enhance the spin injection efficiency due to smaller conductivity mismatch as compared to the HMF/semiconductor heterostructures.[28] Therefore, SGS behaviour with high Curie temperature ($T_C$) makes CFMS thin films potential candidates for spintronics applications (e.g., spin injection).